\documentclass[preprint,12pt]{elsarticle}

\usepackage{amssymb}
\usepackage{amsmath,amsthm,amsfonts}
\usepackage{url}

\usepackage[utf8]{inputenc}
\usepackage[english]{babel}

\usepackage{amssymb}

\usepackage[font=small,labelfont=bf]{caption}

\usepackage[T1]{fontenc}
\usepackage{etoolbox}
\usepackage{cleveref}
\usepackage{multirow}
\usepackage{amsbsy}
\usepackage[top=1.5in, bottom=1.5in, left=0.8in, right=0.8in]{geometry}
\usepackage{color}
\usepackage{tikz}
\usetikzlibrary{patterns}
\usepackage{cleveref}
\usepackage{booktabs}
\usepackage{verbatim}
\usepackage{subfigure}
\usepackage{siunitx}	
\usepackage{float}

\def\({\text{\huge (}}
\def\){\text{\huge )}}

\def\]{\text{\huge ]}}
\def\[{\text{\huge [}}

\newcommand{\beas}{\begin{eqnarray*}}
\newcommand{\eeas}{\end{eqnarray*}}
\newcommand{\bea}{\begin{eqnarray}}
\newcommand{\eea}{\end{eqnarray}}
\newcommand{\be}{\begin{equation}}
\newcommand{\ee}{\end{equation}}

\newcommand{\lb}{\label}

\newcommand{\ra}{\rightarrow}

\newcommand{\pad}[2]{\frac{ \partial #1}{ \partial #2}}

\newcommand{\etal}{\emph{et al}.\ }
\newcommand\nc{\newcommand}
\nc\ord[1]{{\cal O}(#1)}

\DeclareMathOperator{\Le}{Le}	
\DeclareMathOperator{\Sc}{Sc}	
\DeclareMathOperator{\Nt}{Nt}	
\DeclareMathOperator{\Nb}{Nb}	
\DeclareMathOperator{\Nu}{Nu}	
\DeclareMathOperator{\Rey}{Re}	

\begin{document}

\begin{frontmatter}



\title{Does mathematics contribute to the nanofluid debate?}


\author[CRM,UPC]{T.G. Myers}\author[CRM,UPC]{H. Ribera}
\author[CRM]{V. Cregan}
\address[CRM]{Centre de Recerca Matem\`{a}tica, Campus de Bellaterra, Edifici C, 08193 Bellaterra, Barcelona, Spain}
\address[UPC]{Departament de Matem\`{a}tiques, Universitat Polit\`{e}cnica de Catalunya,
              Barcelona, Spain}

\begin{abstract}
Recent experimental evidence has clearly demonstrated that nanofluids do not provide the greatly enhanced heat transfer predicted in the past. Despite seemingly conclusive proof there is still a great deal of current mathematical research asserting  the opposite result. In this paper we scrutinise the mathematical work and demonstrate that the disagreement can be traced to a number of issues. These include the incorrect formulation of the governing equations; the use of parameter values orders of magnitude different to the true values (some requiring nanoparticle volume fractions greater than unity and nanoparticles smaller than atoms); model choices that are based on permitting a reduction using similarity variables as opposed to representing an actual physical situation; presentation of results using different scalings for each fluid.
\end{abstract}


\end{frontmatter}



\section{Introduction}
\label{Sect:Intro}

There exists a wide experimental literature concerning the  heat transfer properties of nanofluids. In the past remarkable increases in thermal conductivity, viscosity and heat transfer coefficient were reported with the addition of a very small volume fraction of nanoparticles  to a base fluid. The experimental work was supported by both theoretical and numerical investigations. However, a remarkable spread in the experimental data  prompted a benchmark study by 34 laboratories around the world \cite{Buon09}. One of their main conclusions was that \emph{no anomalous enhancement
of thermal conductivity was observed in the limited set of nanofluids tested in this exercise.} This result is backed up by other recent studies. The Stanford NanoHeat Group state that \emph{particle based nanofluids show little promise for heat exchanger design}, although they do suggest it is possible carbon nanotube (CNT) based fluids may be viable \cite{StandfordNano}. Putra \etal \cite{Putr03} state that with the addition of nanoparticles \emph{a systematic and definite deterioration in natural convective heat transfer has been found to occur}. Similar results are reported in \cite{Das03, Hagh15, Li10, Wen05}.

Despite  very convincing  experimental evidence that nanofluids are not the predicted saviours in the heat transfer world  there is still a great deal of research activity in this area. One particularly lively area is in the mathematical modelling of nanofluid flow using the system of equations developed by Buongiorno \cite{Buon06}. In particular, practitioners of boundary layer theory, similarity solutions and the Homotopy Analysis Method have published thousands of papers dealing with  different forms of and extensions to Buongiorno's equations and subject to a variety of boundary conditions. In contrast to the recent experimental results these authors are  unanimous in the opinion that nanoparticles have a positive effect on the thermal field and heat transfer characteristics. This is summarised in the review of Haddad \etal \cite{Hadd12} who state that theoretical results show that  \emph{nanofluids significantly improve the heat transfer capability of conventional heat transfer fluids whereas experimental results showed that presence of nanoparticles deteriorates heat transfer systematically.} In \cite{Hadd12b} it is stated that theory shows
\emph{there is always an enhancement in heat transfer by the presence of nanoparticles}.

The mathematical analysis of Buongiorno's equations appears to start with the work of Tzou \cite{Tzou08}
which involves a study of the linear stability of a heated, thin layer of nanofluid.  In that paper. Buorgiorno's equations were extended to include buoyancy in the fluid momentum equation. The results showed a decrease in the critical Rayleigh number of two orders of magnitude. In conclusion Tzou stated  that \emph{clearly, Brownian motion and thermophoresis of nanoparticles introduce additional nonlinear effects for heat transport in nanofluids}. This is despite a subsequent statement that the effect of nanoparticles is only noticeable for a Lewis number  below 10, while accepting that for actual nanofluids the true value is three to four orders of magnitude greater. In fact, as we will see later the value is often even higher.

Tzou's work was followed by a series of papers by Neild and Kuznetsov dealing with nanofluid stability  and convective flow \cite{Kuzn10,Nield2009,Neil09b}. In \cite{Neil09b} they investigate convection driven flow past a vertical plate placed in a porous medium and so write down a modified form of Buongiorno's system to account for buoyancy and  porosity. This is perhaps the first published paper where standard boundary layer theory is applied to the nanofluid equations. In \cite{Kuzn10} they examine the same problem without porosity. Khan and Pop \cite{Khan2010} employ the equations provided in \cite{Kuzn10,Neil09b} but investigate a simpler configuration where flow is driven by the movement of a `stretching surface'. Consequently their system contains no buoyancy terms and in fact more closely follows Buongiorno's original model.

The papers \cite{Kuzn10,Neil09b,Khan2010} use standard boundary layer approximations and similarity variables to reduce the problem to a set of  ordinary differential equations. They have a high number of citations and consequently a plethora of papers follow employing similar reductions and the same basic form of heat and nanoparticle concentration equation. These extensions and modifications  include magnetohydrodynamic effects; radiative heat flux in the heat equation; permeable substrates; heat generation/absorption; non-Newtonian fluids; flow in a cylindrical geometry; flow in a cylindrical geometry embedded in a porous medium; a permeable cone in a porous media; various far-field flow configurations; nanofluids with micro-organisms, see \cite{Bhattacharyya2014,Haya15,ishak2011mhd,Ella13,Raptis2004,Rashad2011,Sheikholeslami2014,Xu14}. Simply for the stretching sheet model there are studies with sheets moving at a constant rate, with velocity proportional to distance $x$; proportional to $x^n$; proportional to $x/t$ (and then with a substrate temperature proportional to $x/t^2$); exponentially increasing \cite{Bhattacharyya2014,Das15,Haya15c,Rana12,Sajid2008}.

Increasing the  model permutations, there is also a wide variety of boundary conditions on temperature (constant, fixed flux, convective, radiative); velocity (no slip, various forms of slip, suction, etc.) \cite{ Alsa12,Hama12,Ibrahim2013,Nogh12,Sheikholeslami2014}. In the absence of a permeable surface the physically sensible boundary condition for the nanoparticles at the substrate is zero flux, i.e., the particles cannot penetrate the substrate. This was imposed in the original paper of Buongiorno \cite{Buon06}. However, the early mathematical studies and many recent ones specify the particle concentration at the substrate \cite{Alsa12,Bach12,Must11,Tham14}.
In \cite{Hadd12b} it is pointed out that this condition is `somewhat arbitrary' and so they employ the zero particle flux condition. Neild and Kuznetsov \cite{Kuznetsov2013,Kuzn14} then use this condition to revisit their previous problems. Tham \etal \cite{Tham14} retain a fixed concentration condition while noting that the zero flux condition or \emph{the `new' boundary condition of \cite{Kuznetsov2013} is a more natural (physical) assumption}.

The physical justification for the mathematical configurations invariably relates to extrusion of polymer sheets and filaments, melt spinning, manufacture of plastic and rubber sheets and cooling of large metallic plates \cite{Khan2010,Nazar2004,Takhar2001}. The actual relation between model and the physical system can be difficult to determine. For example, none of these proposed processes involves an exponentially stretching sheet with a constant temperature moving through a porous medium. In fact the justification for the exponentially stretching sheet studied in \cite{Sanjayanand2006} is simply a comment in \cite{Gupta1977} that the velocity may not be linear: \cite{nadeem2012boundary} then cites \cite{Sanjayanand2006} as a justification. In \cite{Nogh12} slip flow over a stretching sheet is studied. The  slip lengths quoted are up to \SI{68}{mm}, which is an incredibly large number, see \cite{Myer10}. These values come from early experiments on flow in carbon nanotubes \cite{Maj05a}, which is obviously a completely different physical situation. Further, the values are incorrect, in an erratum to the experimental paper, \cite{Maj05b}, it is stated that the slip lengths should have been written as microns, rather than millimetres.  There is also a general avoidance of actual physical parameter values in the mathematical studies, instead  values for the non-dimensional numbers  are taken from previous papers, so propagating errors. However, if they are discussed, the nanofluids are standard, such as water or ethylene glycol containing Cu, CuO, Al$_2$O$_3$, TiO$_2$ or Au nanoparticles, see \cite{Das15,Haya15,Ibrahim2013,Tzou08} for example. In \cite{Raju15} the study focuses specifically on ethylene glycol with Cu or CuO particles, they also use standard relations to determine the nanofluid properties. So the parameter values should be very similar to those in Buongiorno's original paper and in general are easily obtained from the literature.

The enhanced heat transfer predicted by the theoretical studies is in direct contradiction of many experimental results. At the end of their experimental study Li and Peterson\cite{Li10} question the contradiction
\emph{the controversy resulted from simulation study and experimental study is still not clear} and go on to discuss possible mechanisms for this. Since the nanofluids discussed are standard it is clear that the disagreement cannot arise due to the choice of fluid. Consequently, the goal of this paper is to determine the cause of the disagreement. We will show unequivocally that it arises from a series of errors and incorrect values used in the mathematical models. The problems will be illustrated by working through a specific flow configuration common in the literature. However, the conclusions are the same for many other flow configurations.


\section{Mathematical model}
 \label{Sect:Model}

The key parameters in this debate relate to the motion due to Brownian diffusion and thermophoresis.  The Brownian diffusion refers to the effect the Brownian motion of the base fluid molecules has on the nanoparticles.  Thermophoresis (or the Soret effect specifically in liquids) is a related effect whereby the presence of a temperature gradient drives the  motion: hot molecules have more kinetic energy than cool ones, so hot base fluid molecules tend to drive the nanoparticles in the direction of lower temperature. The two terms are quantified by the following expressions
\bea
D_B  = \frac{k_B T }{3\pi \mu_{bf} d_p}\,, \qquad \qquad
D_T = \frac{0.26 k_{bf}}{2k_{bf}+ k_p}\frac{\mu_{bf}}{\rho_{bf}}\phi \, ,
\label{eqn:DefinitionDBDT}
\eea
where $k_B$ is the Boltzmann constant, $T$ is temperature,   $\mu$ the dynamic viscosity, $d_p$ the nanoparticle diameter, $k$ the thermal conductivity, $\rho$ is density and $\phi$ the particle volume fraction. The subscripts $p, bf, nf$ refer to particle, base fluid  and nanofluid, respectively. The above terms are dependent on the local temperature and particle concentration (as opposed to the bulk/far-field values).

The mass flux of nanoparticles due to Brownian motion and thermophoresis is defined as
\bea
\lb{flux}
\mathbf{j} = -\rho_p\left[ D_B \nabla \phi + D_T \frac{\nabla T}{T} \right] \, .
\eea
Since $D_B$ and $D_T$ involve temperature and volume fraction, we apply the notation of   \cite{MacD13} so that the governing equations are formulated in terms of the  `diffusion parameters' $C_B = D_B/T$ and $C_T = D_T/\phi$. This notation permits the variation of $\phi, T$ to be clearly identified in the governing equations. Hence, according to \cite{MacD13}, the governing equations to describe the flow of a heated, Newtonian nanofluid are
\bea
\lb{incom}
\pad{\rho_{nf}}{t} + \nabla \cdot (\rho_{nf}\mathbf{u}) = 0 \,,\\
\lb{ueq}
\rho_{nf}   \left[ \pad{\mathbf{u}}{t} + \mathbf{u}\cdot\nabla   \mathbf{u} \right]  = - \nabla p - \nabla \cdot \boldsymbol\tau \,,
\\
\lb{Teq}
\pad{(\rho c)_{nf} T}{t} + \nabla\cdot( (\rho c)_{nf} \mathbf{u}  T )  = \nabla\cdot( k_{nf} \nabla  T ) + \rho_p c_p
\left[ C_B T \nabla \phi + C_T \phi \frac{\nabla T}{T} \right] \cdot \nabla T \,,
\\
\lb{phieq}
\pad{\phi}{t}  + \mathbf{u}\cdot\nabla  \phi  = \nabla \cdot \left[ C_B T \nabla \phi + C_T \phi \frac{\nabla T}{T} \right] \, ,
\eea
where $\mathbf{u}$ is the nanofluid velocity, $p$ is pressure, $\boldsymbol\tau$ is the stress tensor and $(\rho c)$ is the  volumetric heat capacity. Gravity and viscous dissipation have been  assumed negligible. This system differs slightly from that presented in \cite{Buon06}. In \cite{Buon06}  the fluid is assumed incompressible, although in reality the nanofluid density varies linearly with volume fraction.  In fact, the properties of the nanofluid $\rho_{nf}, k_{nf}, (\rho c)_{nf}, \mu_{nf}$ all depend strongly on the particle concentration. Indeed, it is this dependence that has generated a lot of the excitement around nanofluids. Thus, to reflect its variation with respect to $\phi$, the density, $\rho_{nf}$, and volumetric heat capacity $(\rho c)_{nf}$ have been kept inside the derivatives.  A strict derivation of (\ref{incom}-\ref{phieq}), valid for compressible and incompressible nanofluids, is given in \cite[\S 5.7.3]{MacD14}.


\section{Reduced governing equations}

We will now reduce the governing equations in accordance with standard boundary layer theory. To do this we will focus on one of the simplest  possible configuration used in the literature, which allows us to demonstrate the typical form of solution  without making the analysis overly complicated.
To this end we specify a Newtonian fluid flow driven by a thin stretching sheet surrounded on either side by fluid. The sheet has a higher temperature than the surrounding fluid. This temperature remains unaffected by the fluid and is transmitted infinitely well, so that the fluid immediately in contact with the sheet has exactly the same temperature. The configuration is depicted in \cite{Khan2010}. To avoid confusion we note that \cite{Khan2010} incorrectly identifies the coefficient $\rho_{nf}$ in equation \eqref{ueq} as the base fluid density, this follows from the work of \cite{Kuzn10,Nield2009} who ambiguously define it as the fluid density (and the error is propagated in  \cite{Maki11} who then use the base fluid density in the nanofluid momentum equation and the base fluid  volumetric heat in the nanofluid energy equation).

The only difference we make with the model of \cite{Khan2010} is with the boundary conditions, rather than fixing the  particle concentration at the wall we impose  the zero flux condition. Considering the  surface of the sheet to be at $y=0$ we therefore specify
\bea
\lb{bc1}
u&=& a x \, , \quad v=0 \, , \quad T=T_w  \, ,\quad C_B T \nabla \phi + C_T \phi \frac{\nabla T}{T} = 0 \quad \mbox{at} \, \, \, y=0 \, ,\\
\lb{bc2}
u&=& v=0  \, , \quad T=T_{\infty}  \, , \quad \phi = \phi_{\infty} \quad \mbox{as} \, \, \, y\rightarrow \infty \, .
\eea
Obviously even with the zero particle flux condition this is a rather unrealistic set of conditions. The surface $y=0$ is stretching with a speed proportional to position, $u=ax$, yet there is no vertical displacement of the surface to account for  mass conservation. Moreover,  the thin sheet acts to heat the fluid, yet retains a constant temperature throughout the process. However, we will leave the conditions as stated since the solution method employed in the analysis of these equations usually requires unphysical conditions.

The governing equations will now be reduced in line with standard boundary layer theory. This has been used to solve some classical problems and starts from a set of physically sensible assumptions. For example, the flow is steady; the thickness of the boundary layer, where rapid changes occur in the flow, is narrow compared to the length-scale of the flow; the velocity parallel to the wall is much higher than the perpendicular velocity; the derivatives perpendicular to the wall are large (since rapid change occurs over a thin region), see \cite{Acheson, white2006viscous}. Under these assumptions the nanofluid equations may be written
\bea
\lb{incombl}
 \nabla \cdot (\rho_{nf}\mathbf{u}) &=&  0 \,,\\
\lb{phibl}
\mathbf{u}\cdot\nabla  \phi  &= &\pad{}{y} \left[ C_B T \pad{\phi}{y} + C_T \frac{\phi}{T} \pad T y\right] \,, \\
\lb{Tbl}
\nabla\cdot( (\rho c)_{nf} \mathbf{u}  T )  &=& \pad{}{y}\left( k_{nf} \pad T y \right) + \rho_p c_p \left[C_B T \pad{\phi}{y} \pad T y + C_T \frac{\phi}{T} \left(\pad T y\right)^2 \right] \, ,\\
\lb{ubl}
\rho_{nf}   \mathbf{u}\cdot\nabla   u     &=&  - \pad p x -   \pad{}{y} \left(\mu_{nf} \pad u y\right)   \quad \,, \qquad -\pad p y = 0 \, .
\eea
The final equation indicates $p=p(x)$. If we assume a constant far-field pressure (the flow in this case is driven by the wall motion), $\left. p \right|_{y \ra \infty}= p_{\infty}$,  then the pressure must be constant everywhere and so $p_x=0$.

As stated, the nanofluid properties depend on the volume fraction:  density and volumetric heat capacity may be well approximated by simple linear laws while
viscosity and thermal conductivity vary nonlinearly and the exact relations are still subject to debate, see \cite{MacD13} and references therein. However, if we wish to follow the approach of previous analyses we must now neglect this dependence and set the thermophysical parameters to their bulk values, $(\rho c)_{\infty}, k_{\infty}, \rho_{\infty}, \mu_{\infty}$, where for example $\mu_{\infty}=\mu_{nf}(\phi_\infty)$. In the literature this is not justified via any linear theory, rather the equations are written from the start with the parameters outside the derivatives, see \cite{Hami15,Khan15,Khan2010, Kuzn10,Kuzn14,Neil09b,Tham14,Tzou08} for example.

The model reduction proceeds via the following change of variables
\bea
\lb{nondim}
\theta(\eta)= \frac{T(x,y)-T_{\infty}}{\Delta T} \,, \qquad
\chi(\eta) = \frac{\phi(x,y)-\phi_{\infty}}{\phi_{\infty}} \, ,
\eea
where $\Delta T = T_w-T_{\infty}$.
The velocity is written in terms of the stream function
\bea
\lb{uvpsi}
u = \pad{\psi}{y} \,,
\qquad v = -\pad{\psi}{x} \,,
\eea
where
\bea
\lb{psi}
\psi = \left( a \nu_{\infty}\right)^{1/2} x f(\eta) \,, \qquad \eta = \left(\frac{a }{\nu_{\infty}}\right)^{1/2}
 y \, ,
\eea
and $\nu_{\infty}=\mu_{\infty}/\rho_{\infty}$, see \cite{Khan2010}.
Substituting these new variables into (\ref{phibl}--\ref{ubl}), with constant values for the thermophysical properties, leads to
\begin{align}
\lb{chieq}
\chi''+ \frac{\Delta T}{T_{\infty}} \theta' \chi' + \frac{\Nt}{\Nb}  \left[\theta'\chi' + (1+\chi) \theta'' -  \frac{\Delta T}{  T_{\infty}} (1+\chi) \theta'^{2}\right] + \Sc f \chi' &= 0 \,,\\
\lb{theq}\frac{1}{\Pr} \theta'' + f \theta' + \Nb \chi'\theta' + \Nt (1+\chi) \theta^{'2} &= 0\,,\\
\lb{feq} f''' + f f'' -f^{'2}&= 0 \,,
\end{align}
where primes denote differentiation with respect to $\eta$. In deriving (\ref{chieq}, \ref{theq}) we have followed the linearisation suggested in \cite{Buon06}, which states $\Delta T \ll T_{\infty}$. Obviously this places some restriction on the permissible temperature fields but it allows us to replace $\Delta T \,  \theta + T_{\infty}$ by $T_{\infty}$. This has been carried out in writing down the denominators in two terms of (\ref{chieq}). However, since we are investigating boundary layer flow, where gradients may be high, we have retained terms involving $\Delta T/T_{\infty}$ whenever they multiply a derivative. If the gradients are not high then these terms will be small and so not affect the results.

The dimensionless parameters are defined  to be
\begin{equation}
\label{LSPDef}
\Sc = \frac{\nu_{\infty}}{  C_B T_{\infty}}, \qquad
\Pr = \frac{\nu_{\infty}}{\alpha_{\infty}} , \qquad
\Nb = \frac{ C_B T_{\infty} \phi_{\infty}  }{  \nu_{\infty}}\frac{\rho_p c_p }{(\rho  c)_{\infty} }, \qquad
\Nt = \frac{  C_T \phi_{\infty} \Delta T  }{  \nu_{\infty} T_{\infty}}  \frac{\rho_p c_p }{(\rho  c)_{\infty} } \, .
\end{equation}
The coefficient $\Sc$ is the Schmidt number, and it represents the ratio of viscous to Brownian diffusion. In studies where the flow is driven by buoyancy, the Schmidt number must be replaced by the Lewis number, $\Le=\alpha_{\infty}/C_B T_{\infty}$, the ratio of thermal to Brownian diffusion. The two numbers are often confused in the literature, see \cite{Abba15, Alsa12, Bach12,  Hami15,  Hass11, Khan2010,  Maki11, Rana12} for example. The Prandtl number, $\Pr$, is the ratio of viscous to thermal diffusivity.  The numbers $\Nb, \Nt$ represent the ratios of Brownian and thermpohoretic diffusion to viscous diffusion while the ratio $\Nt/\Nb =  C_T \Delta T/(C_B T_{\infty}^2)$ represents the relative strength of thermophoretic to Brownian diffusion.

The boundary conditions (\ref{bc1},\ref{bc2}) transform to
\bea
\lb{bceta1}
f(0)=0\,, \qquad f'(0)=1 \,,\qquad \theta(0) = 1 \,,\qquad  \chi'(0) + \frac{\Nt}{\Nb} (1+\chi(0))\, \theta'(0) = 0 \, ,
\\
\lb{bceta2}
f(\infty)=0 \,,\qquad f'(\infty) = 0\,, \qquad \theta(\infty)=0\,, \qquad \chi(\infty) = 0 \, .
\eea

\subsection{Previous governing equations}

The above set of governing equations, (\ref{chieq}-\ref{feq}),  is markedly different to those studied previously in the literature. In \cite{Hami15,Hass11,Khan2010,Maki11,Must11,Must13,Nogh12,Rana12} the transformation of the nanoparticle concentration and heat equations is given as
\bea
\label{Tzouchi}
\chi'' + \frac{\Nt}{\Nb} \theta'' +  \Sc f \chi' &=& 0 \,, \\
\label{Tzouth}
\frac{1}{\Pr} \theta'' + f \theta' + \Nb \chi'\theta' + \Nt \theta'^{2} &=&  0\, .
\eea
 In some of these works the velocity equation differs  from \eqref{feq} due to a different stretching mechanism (such as \lq nonlinear' or exponential). They all incorrectly denote the  Schmidt number  as the Lewis number.

Going back to the original analysis of Tzou \cite{Tzou08} we can trace  differences to the use of `linearisation'.  Tzou \cite{Tzou08} states that
$\Delta T \ll T_{\infty}$ and so replaces the denominator $T$ by $T_{\infty}$ in the flux expression of equation \eqref{flux}, he also assumes that the coefficients $D_B, D_T$ are constant. The same assumptions are stated by Buongiornoin in \cite{Buon06}. The replacement of $T$ by $T_{\infty}$ in the flux  is similar to our replacement of $T$ in deriving \eqref{chieq} but we have allowed for the possibility of large gradients and so retained the $T$ derivative resulting from the $T \phi_y$ term. The coefficient $D_T = C_T \phi$ contains $\phi$, and taking it as constant is equivalent to assuming that the $\phi$ variation is small. There is no justification for this, particularly since the interesting fluid behaviour occurs when $\phi$ varies.

From this discussion we see that equations (\ref{Tzouchi}, \ref{Tzouth}) may be derived from (\ref{phibl}, \ref{Tbl}) by setting every appearance of $T, \phi$ that are not in a derivative to $T_{\infty}, \phi_{\infty}$. This results in the loss of three terms from \eqref{chieq} as well as the replacement of  $1+\chi$ by $1$ in equations (\ref{chieq},\ref{theq}) (since $\phi=\phi_{\infty}$ implies $\chi=0$). It also affects the no-particle flux condition in  \eqref{bceta1}.

The final governing equation, \eqref{feq}, is third order in $\eta$ and is solved exactly in the studies of flow driven by a linear stretching sheet \cite{Hass11, Khan2010,Maki11} after applying the boundary conditions $f(0)=0, f'(0)=1, f'(\infty)=0$. The quoted result is
\bea
f = 1- \exp(-\eta) \, .
\lb{eqn:Solutionf}
\
\eea

There is a minor issue with this solution which is not discussed in previous papers. Using a standard similarity variable $\eta \propto y/x^n$, with $n > 0$ the conditions in the far-field coincide with those at the inlet, since $\eta = \infty$ when either $y=\infty$ or $x=0$. This reduces the number of boundary conditions in the $\eta$ domain (and also restricts the permissible flow configurations, since the inlet and far-field conditions must match). For the present case $\eta \propto y$ and there is no reduction in the number of boundary conditions. This is dealt with in the past studies by simply neglecting the condition $\left. v \right|_{y \ra \infty} = 0$. Using the definitions (\ref{uvpsi}, \ref{psi}) and solution (\ref{eqn:Solutionf}) gives $\left. v \right|_{y \ra \infty} = -(a \nu_{\infty})^{1/2} \left. f \right|_{\eta \ra \infty} = -(a \nu_{\infty})^{1/2}$, which violates the far-field condition. However, since $\nu$ is small the error is also small. While this error is not common to all flow configurations it does appear in the linear stretching sheet literature. A similar problem may be found with the exponentially stretching sheet configuration of \cite{Must15}.

\section{Parameter values}

We have now presented the equations governing the boundary layer flow and highlighted problems with the previous formulations. Before going on to solve the system we must first specify appropriate values for the non-dimensional numbers $\Sc, \Pr, \Nt, \Nb$. We will also discuss the Lewis number, $\Le$, since this appears in studies where buoyancy drives the flow. To calculate the non-dimensional numbers requires information concerning the nanofluids. In Table \ref{PropVals} we provide values for the density, specific heat and conductivity of five standard base fluids and four nanoparticles. We also provide the viscosity values for the fluids.

\begin{table}[H]
\centering
\begin{tabular}{lcccc}
\hline
\multirow{2}{*}{Material} & $\rho$ & $c$ & $k$ & $\mu$  \\
 & (kg/m$^3$) & (J/kg K) & (W/m K) & (Pa s) \\ \specialrule{1.3pt}{1pt}{1pt}

Water  & 998 & 4190 & 0.58 & 0.001  \\ \hline

Ethylene glycol (EG) & 1110 & 2470 & 0.258 & 0.014 \\ \hline

Engine oil SAE 30 (EO) & 878 & 2090 & 0.145 &  0.0153 \\ \hline

Polyvinylpyrrolidone (PVP)  & 1200 & 3349 & 0.15 & 0.012 \\ \hline

Polyalphaolefin (PAO) & 814 & 2135 & 0.159 & 0.0161 \\ \hline

Alumina (Al$_2$O$_3$) & 3690 & 880 & 35 & - \\ \hline

Copper (Cu) & 8960 & 386 & 385  & - \\ \hline

Gold (Au) & 19300 & 129 & 318 & - \\ \hline

Titanium dioxide (TiO$_2$) & 4230 & 683 & 11.8 & - \\ \hline

\end{tabular}
\caption{Thermophysical  parameters for common base fluids and nanoparticles.  The data for the fluids corresponds to water at \SI{293}{K}, EG at \SI{300}{K}, EO at \SI{353}{K}, PVP at \SI{296}{K} and PAO at \SI{313}{K}.}
\label{PropVals}
\end{table}

In Table \ref{LSPTable} a range of Lewis, Schmidt and Prandtl numbers are provided for typical nanofluids using data from Table \ref{PropVals}. To calculate $\Sc, \Le$ we  also require the Boltzmann constant $k_B \approx 1.38 \times 10^{-23}$ m$^2$ kg s$^{-2}$ K$^{-1}$.

The minimum and maximum values of the Lewis and Schmidt number correspond to particles with diameter $d_p=$1 to \SI{100}{nm} and a far-field concentration $\phi=0.01$. The thermal conductivity of the nanofluid  was calculated using the formula given in \cite{Myers2013}. The density and the volumetric heat capacity were calculated using the standard linear laws
\bea
\rho_{nf} &=& \phi \rho_{p} + (1-\phi) \rho_{bf} \lb{rhoeq} \,,  \\
(\rho c)_{nf} &=& \phi \rho_p c_p + (1-\phi) \rho_{bf} c_{bf} \, . \lb{rhoceq}
\eea

\begin{table}[H]
\centering
\begin{tabular}{lccccc}
\hline
 Nanofluid & $\min(\Le)$ & $\max(\Le)$ & $\min(\Sc)$ & $\max(\Sc)$ & $\Pr$ \\ \specialrule{1.3pt}{1pt}{1pt}

Water/Al$_2$O$_3$ & $4.32 \times 10^{4}$ & $4.36 \times 10^{6}$ & $2.28 \times 10^{3}$ & $2.3 \times 10^{5}$ & $7.24$ \\ \hline

Water/Cu & $4.05 \times 10^{4}$ & $4.1 \times 10^{6}$ & $2.17 \times 10^{3}$ & $2.19 \times 10^{5}$ & $7.24$ \\ \hline

EG/Al$_2$O$_3$ & $3.1 \times 10^{5}$ & $3.14 \times 10^{7}$ & $4.02 \times 10^{5}$ & $4.06 \times 10^{7}$ & $134$ \\ \hline

Engine oil/TiO$_2$ & $1.55 \times 10^{5}$ & $1.56 \times 10^{7}$ & $4.95 \times 10^{5}$ & $5 \times 10^{7}$ & $220$ \\ \hline

PVP/Cu & $1.24 \times 10^{5}$ & $1.26 \times 10^{7}$ & $2.6 \times 10^{5}$ & $2.62\times 10^{7}$ & $268$ \\ \hline

PVP/Au & $1.72 \times 10^{5}$ & $1.74 \times 10^{7}$ & $2.4 \times 10^{5}$ & $2.43\times 10^{7}$ & $268$ \\ \hline

PAO/Al$_2$O$_3$ & $2.12 \times 10^{5}$ & $2.14 \times 10^{7}$ & $6.92 \times 10^{5}$ & $6.99 \times 10^{7}$ & $216$ \\ \hline

\end{tabular}
\caption{Values of Lewis, Schmidt and Prandtl numbers for  $d_p =$ 1 - \SI{100}{nm} and $\phi=0.01$. The temperatures are water at \SI{293}{K}, EG at \SI{300}{K}, EO at \SI{353}{K}, PVP at \SI{296}{K} and PAO at \SI{313}{K}.}
\label{LSPTable}
\end{table}

In Table \ref{NbtTable} we present values for the Brownian motion and thermophoresis parameters using the same range of particle sizes, volume fraction and temperature as in the previous tables. We also require a value of  $\Delta T$ for $\Nt$ and so choose $\Delta T =$ \SI{10}{K}. For water this leads to values of $\Nb$ in the range $[10^{-8}, 10^{-6}]$ and $\Nt$ of order $10^{-6}$. The remaining fluids show $\Nb$ ranging from $[10^{-10}, 10^{-8}]$ and $\Nt$ of orders between $[10^{-8}, 10^{-6}]$.
\begin{table}[h]
\centering
\begin{tabular}{lccc}
\hline

Nanofluid & $\min(\Nb)$ & $\max(\Nb)$ & $\Nt$   \\ \specialrule{1.3pt}{1pt}{1pt}

Water/Al$_2$O$_3$ & $3.37 \times 10^{-8}$ & $3.4 \times 10^{-6}$ & $1.1 \times 10^{-6}$ \\ \hline

Water/Cu & $3.77 \times 10^{-8}$ & $3.81 \times 10^{-6}$ & $1.1 \times 10^{-6}$\\ \hline

EG/Al$_2$O$_3$ & $2.47 \times 10^{-10}$ & $2.5 \times 10^{-8}$ & $6.32 \times 10^{-7}$ \\ \hline

Engine oil/TiO$_2$ & $3.15 \times 10^{-10}$ & $3.18 \times 10^{-8}$ & $1.39 \times 10^{-6}$ \\ \hline

PVP/Cu & $3.28 \times 10^{-10}$ & $3.31 \times 10^{-8}$ & $2.94 \times 10^{-8}$ \\ \hline

PVP/Au & $2.55 \times 10^{-10}$ & $2.58 \times 10^{-8}$ & $2.56 \times 10^{-8}$\\ \hline

PAO/Al$_2$O$_3$ & $2.26 \times 10^{-10}$ & $2.29 \times 10^{-8}$ & $5.91 \times 10^{-7}$ \\ \hline

\end{tabular}
\caption{Brownian motion and thermophoresis parameter values. The range of nanoparticle size is $d_p = 1$ nm to $d_p =$ \SI{100}{nm}, the  volume fraction $\phi = 0.01$ and $\Delta T =$  \SI{10}{K}. }
\label{NbtTable}
\end{table}

If we wish to apply the results of the analysis and so make recommendations for future nanofluid design it is useful to be clear exactly what  these parameter values signify. The Prandtl number $\Pr$ is the ratio of momentum to thermal diffusion: a large Prandtl number indicates that energy moves primarily due to the fluid motion, rather than thermal diffusion. The values in Table \ref{LSPTable} vary between 7 for water-alumina to 270 for PVP-gold. The  Schmidt number is the ratio of momentum to Brownian diffusion: the lowest value we calculate is $2  \times 10^{3}$ which corresponds to water with 1 nm alumina particles (and note in practice the smallest particles would be closer to \SI{10}{nm}). In general the value is of order $10^{5}- 10^{7}$. The Lewis number is the ratio of thermal to Brownian diffusion, the lowest value in the Table is $4  \times 10^{4}$ (again for  water-alumina) while the general range is  $10^{5}- 10^{7}$.  The very high values of $\Sc$ and $\Le$ indicate Brownian diffusion plays a very minor role. The Brownian motion parameter $\Nb$ is approximately the ratio of Brownian to viscous diffusion and so is a form of inverse Schmidt number with a correction for the volumetric heat ratio
\bea
\lb{Nbeq}
\Nb = \frac{ \phi_{\infty}  }{ \Sc}\frac{\rho_p c_p }{(\rho  c)_{\infty} }\, .
\eea
The thermophoresis parameter $\Nt$ is approximately the ratio of thermophoretic to viscous diffusion. For a moving liquid we should expect transfer to be primarily via fluid motion, rather than thermophoresis or Brownian motion, and consequently $\Nb, \Nt$ should be small. As we can see from Table \ref{NbtTable} they range between $[10^{-10}, 10^{-6}]$. From this  we may conclude Brownian motion and thermophoresis are not important. Buongiorno \cite{Buon09} provides typical values, $\Le =8 \times 10^5$, $7\times 10^5$ for water-alumina and water-copper nanofluids respectively and gives the ratio $\Nb/\Nt = 0.2, 2$, all of which fit in with the ranges given in the tables.

If we set $\Nb=\Nt=0$ then the governing equations reduce to those of a standard fluid. This can provide a check on the validity of solutions by comparing the limit $\Nb, \Nt \ra 0$ with existing results for the base fluids.

The mathematical studies on stretching sheet flow \cite{Hass11,Khan2010,Must11,Maki11,Nogh12} typically use parameters  of order $\Pr=10$, $\Sc=10$, $\Nb=0.1$, $\Nt=0.1$ (note  all except \cite{Must11} denote $\Sc$ as $\Le$). Lewis numbers between 0.5 and 25 are employed  in \cite{Haya15,  Ibrahim2013, Khan2010,Kuzn10,Kuznetsov2011, Nield2009,Sheikholeslami2014}. Since these are very different to the ranges given in the tables we will now consider their physical meaning.
Firstly, the value $\Pr=10$ indicates  a fluid with properties similar to water. The Schmidt number contains various values that are fixed by the choice of material, such as the density, viscosity etc, however the temperature, volume fraction and particle diameter may be adjusted to vary $\Sc$ and so produce a desired value. Let us take the example of a water based nanofluid  containing 1\% alumina particles ($\phi=0.01$). The definition of the Schmidt number is
\bea
\label{Sceq}
\Sc = \frac{\nu_{\infty}}{C_B T_{\infty}} = \frac{\mu_{\infty} 3 \pi \mu_{bf} d_p}{\rho_{\infty} k_B T_{\infty}} \, .
\eea
The nanofluid density, $\rho_{\infty}= \left. \rho\right|_{\phi=0.01} \approx 1025$ kg/m$^3$, may be determined via equation \eqref{rhoeq}. Maiga \etal \cite{Maig04}  fitted experimental data for water based nanofluids to find a relation for viscosity
\bea
\mu_{nf} = (1+7.3 \phi +123 \phi^2) \mu_{bf} \, .
\eea
For  1\% alumina this shows $\mu_{\infty} \approx 1.085 \mu_{bf}$.
If we wish to determine the largest particles that correspond to a given $\Sc$ then we must take the largest possible value for $T_{\infty}$. For liquid water we will assume $T_{\infty}=373$ K. Equation \eqref{Sceq} then indicates that to obtain $\Sc=10$ we require nanoparticles with diameter $d_p \approx 5 \times 10^{-12}$ m $= 5 \times 10^{-3}$ nm: this is around 1/60th the size of a water molecule. Raju \etal \cite{Raju15} study ethylene glycol based nanofluids containing a very large amount of Cu or CuO particles, $\phi \in [0.1, 0.3]$ or a 10-30\% volume fraction (for such high volume fractions we would expect  non-Newtonian behaviour), and define the linear laws (\ref{rhoeq}, \ref{rhoceq}) for the density and volumetric heat capacity. The viscosity is defined through the Brinkman relation $\mu_{nf} = \mu_{bf}/(1-\phi)^{2.5}$, which is usually quoted as being valid for  low volume fractions $\phi < 0.04$. They have calculations with $\Sc=0.6$. Taking the mean value $\phi=0.2$, $T_{\infty}=300$ K, and other values from Table \ref{PropVals}   indicates that for  Cu  the particle diameter is of the order 10$^{-15}$m (in fact approximately the size of the classical electron). The same order of magnitude holds for CuO particles.

The dependence of density on volume fraction is given by equation \eqref{rhoceq}.
The Brownian motion  parameter $\Nb$ is given by equation \eqref{Nbeq}. Rearranging \eqref{Nbeq} we may express the nanofluid volumetric heat capacity in terms of the product $\Nb \Sc$ and then substituting this into equation \eqref{rhoceq} leads to
\bea
\label{NbSc}
\left[\frac{1}{\Nb \Sc}-1\right] \phi_{\infty} \rho_p c_p    =    (1-\phi_{\infty}) \rho_{bf} c_{bf} \, .
\eea
Nanofluids are suspensions of small quantities of nanoparticles in a base fluid, so the volume fraction of particles is small, in practice typically $\phi < 0.05$. Theoretically we must constrain $\phi \le 1$, where the value $\phi=1$ signifies that the material is now composed entirely of solid particles and no fluid.
If $\phi_{\infty} \le  1$, then the right hand side of equation \eqref{NbSc} is non-negative. To ensure the left hand side is also non-negative requires  $\Nb \Sc \le  1$ and in reality we will have a small volume fraction and so expect $\Nb \Sc \ll 1$. The graphs presented in the study of flow driven by a stretching sheet \cite{Khan2010} have $\Nb \Sc$ in the range $[1,12.5]$: similar values are used in  \cite{Hass11,Must11,Maki11,Nogh12}. Any study that uses the linear relation for volumetric heat capacity with $\Nb \Sc \ge 1$ requires a volume fraction $\phi \ge 1$ and so makes no physical sense. A similar issue arises with the thermophoresis parameter $\Nt$. Even taking a high value of $\Delta T/T_{\infty}= 100/373$ the choice of
$\Nt=0.1$ (used for example in \cite{Anbuchezhian2012,Khan2010,Kuzn10,Mahmoodi2015,Maki11,Nield2009,Nield2011})  requires a volume fraction $\phi \approx 4.47$ for a water-alumina nanofluid.

\section{Model comparison}
\label{Sect:Comparison}

Before comparing the results from the different models we will first discuss the effect of the parameter values on the governing equations. The non-dimensionalisation carried out means that all variables should be order unity. The relative size of the terms in the equations is reflected in the size of the coefficients. Given that the Schmidt number ranges between $2  \times 10^{3}$ to $7  \times 10^{7}$ for all nanofluids given in Table
\ref{LSPTable} we may approximate the current nanoparticle concentration equation,  \eqref{chieq}, or the equation of previous studies,  \eqref{Tzouchi}, by
\bea
f \chi' = 0 \,,
\eea
to an accuracy at worst of the order 0.1\%, when $\Sc=2  \times 10^{3}$, going down to order $10^{-6}$\% when $\Sc=7  \times 10^{7}$. Since we impose $f'(0)=1$, the only possible solution to this equation is $\chi'=0$. Applying the boundary condition at infinity determines $\chi=0$ (and hence $\phi=\phi_{\infty}$ is constant). The term $f \chi' $ is derived from  $\mathbf{u} \cdot \nabla \phi$ in the original equations. The physical meaning of $\mathbf{u} \cdot \nabla \phi = 0$ is that the nanoparticles simply follow the fluid streamlines: there is no noticeable diffusion of particles.

Values of $\Nb, \Nt$ are in the range $2  \times 10^{-10}$ to $4  \times 10^{-6}$, reducing the heat equation to
\bea
\lb{thred}
\frac{1}{\Pr} \theta'' + f \theta' = 0 \, ,
\eea
with errors of the order $10^{-8}$ to $10^{-4}$\%. Note, if the temperature change in the fluid is a mere \SI{10}{K} and the far-field temperature is \SI{300}{K} then the neglect of terms $\Delta T/T$ leads to errors of the order 3\%. An error of $\ord{10^{-4}}$\% therefore seems acceptable given the approximations already made. The physical meaning of this is that Brownian motion and thermophoresis are negligible in comparison to the advection and standard diffusion of heat and so the heat flow is accurately described by the standard model for a particle free fluid and the nanoparticle motion has no effect on it.
This verifies a statement in \cite{Buon06}, based on observation of the non-dimensional system,  \emph{that heat transfer associated with nanoparticle dispersion is negligible compared with  heat conduction and convection. It follows that, in solving nanofluid heat transfer problems ... the energy equation for a nanofluid (is) formally identical to that of a pure fluid}.
In fact the only effect of nanoparticles in either of these equations is in the value of the coefficient $\Pr$ which depends on the viscosity and thermal diffusivity, which in turn depend on the (constant) nanoparticle concentration. This result was previously shown in the study of flow over a flat plate  \cite{MacD13}. In that paper it was demonstrated that the application of Buongiorno's model, at least to water or ethylene glycol with alumina or CuO particles, does not increase the heat transfer coefficient (in fact their results showed a decrease with increasing nanoparticle concentration).

The result of this discussion is that   the flow of a nanofluid over a linearly stretching sheet may be described solely by equation \eqref{thred}, with $f=1-e^{-\eta}$ (whilst noting that the solution for $f$ does not satisfy all velocity boundary conditions). Once appropriate values are applied similar reductions will hold for all the other nanofluid analyses.

\begin{figure}[H]
\centering
\subfigure[Temperature in the boundary layer]{\includegraphics[scale=0.3]{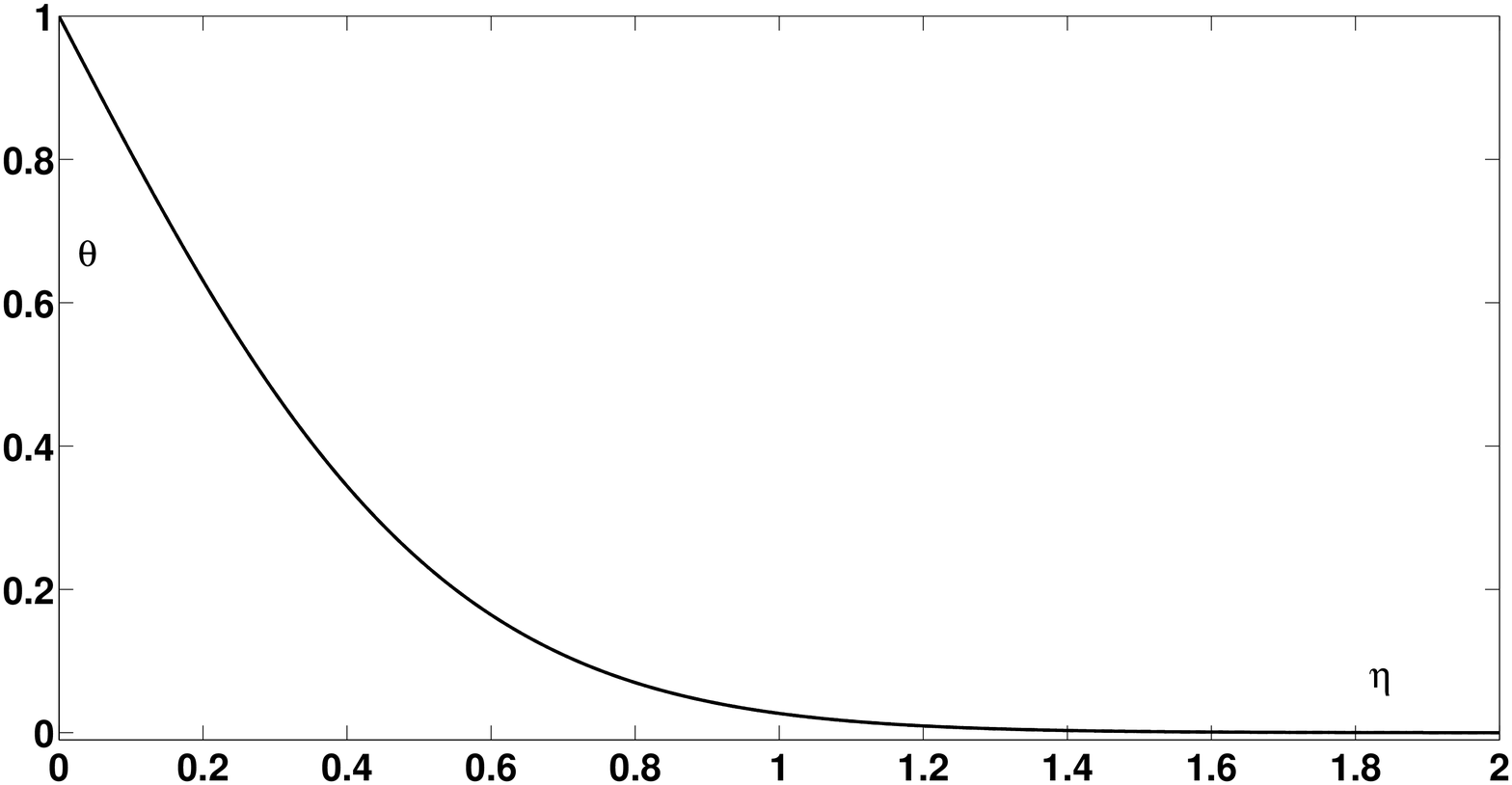}}
\hspace{0.4cm}
\subfigure[Particle concentration in the boundary layer]{\includegraphics[scale=0.3]{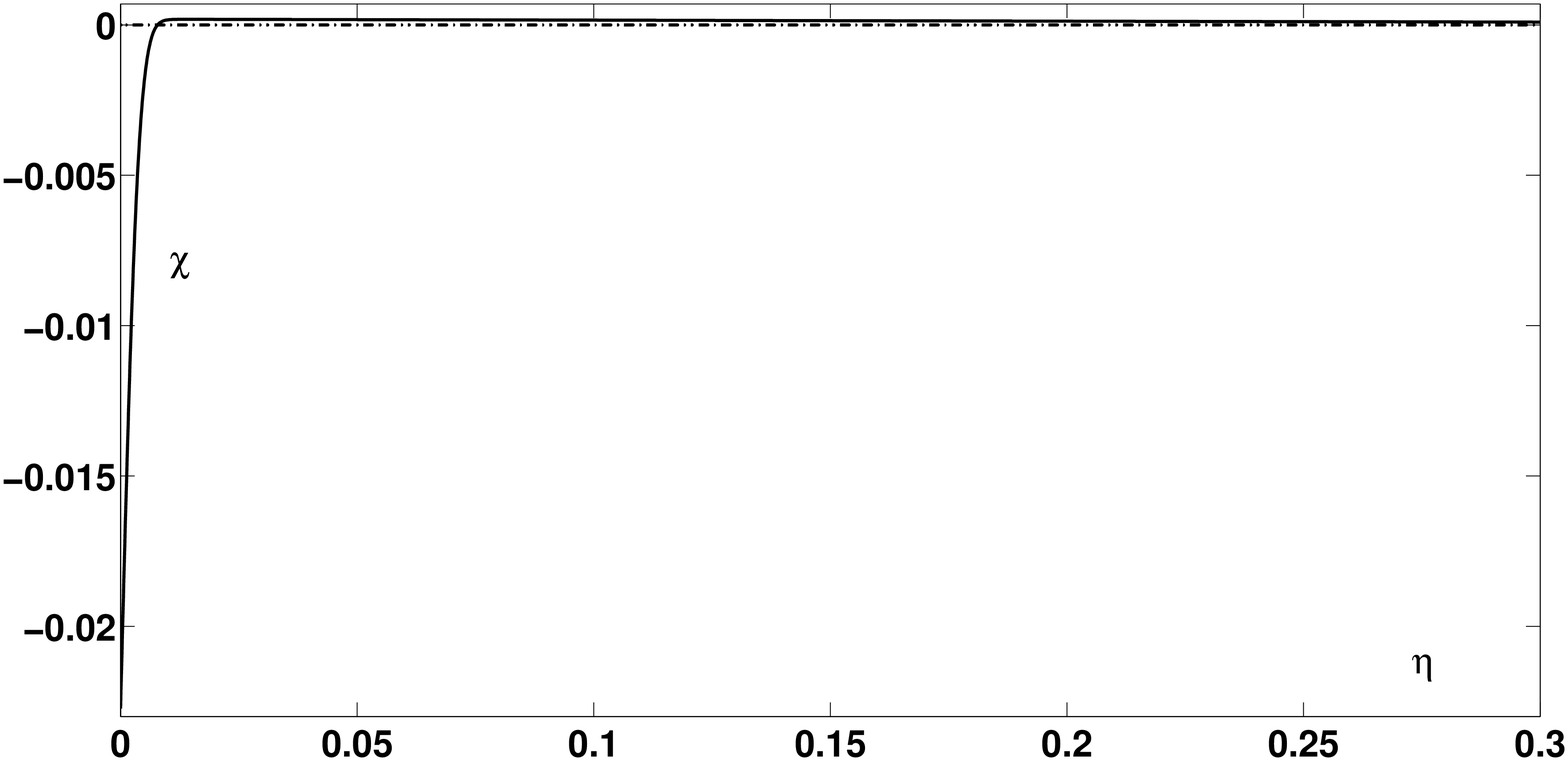}}
\caption{Comparison of a) temperatures and b) particle concentrations predicted by the current model, equations (\ref{chieq}-\ref{feq}) (solid line), and the reduced system of equation \eqref{thred} (dash-dotted line) with $f=1-e^{-\eta}$. Parameter values for a water based nanofluid: $\Sc= 1.16\times 10^5$, $\Pr=7.24$, $\Nb=3.38\times 10^{-7}$, $\Nt = 1.1 \times 10^{-6}$.}
\label{Fig1}
\end{figure}

To verify our statements, in Figure \ref{Fig1} we compare the results of the full model, equations (\ref{chieq}-\ref{feq}), and the reduced system consisting solely of equation \eqref{thred} with $f=1-e^{-\eta}$, subject to conditions (\ref{bceta1}, \ref{bceta2}). The parameters used in the full model have typical values for a water based nanofluid $\Sc= 1.16\times 10^5$, $\Pr=7.24$, $\Nb=3.38\times 10^{-7}$, $\Nt = 1.1 \times 10^{-6}$. In calculating these we have used values $T_w =$ \SI{310}{K}, $T_{\infty}=$\SI{300}{K}, $\Sc, \Nb$ are the mean values of those given in Table \ref{LSPTable}.  Two curves are shown in each figure, the solid line is the solution of the full model, the dash-dot line that of the reduced model. The curves are shown up to a maximum value of $\eta=2$, however to ensure the numerical calculations were past the boundary layer the calculations were in fact carried out to $\eta=10$ (and tested for convergence by using even higher values).  Figure \ref{Fig1}a) shows the temperature profile predicted by the two models, the lines are indistinguishable (in fact the maximum difference is below 0.002\%). If we define the  boundary layer as the region where the value is more than 1\% different from the far-field value, then for the temperature the boundary layer ends at  $\eta \approx 1.2$. If the stretching rate $a=$\SI{1}{cm/s} then this corresponds to a distance $y \approx$ \SI{1.2}{cm}, that is, the thermal boundary layer is approximately \SI{0.012}{m} thick.
Figure \ref{Fig1}b) shows the particle concentrations, note the reduced model predicts $\chi=0$ everywhere. The full model shows a maximum $\chi \approx 1.9 \times 10^{-4}$ and a minimum at the wall of -0.0228 (which corresponds to 0.9772$\phi_{\infty}$). In this case the boundary layer ends at $\eta \approx 2.26 \times 10^{-3}$. This signifies that there is very little movement of nanoparticles and then only in an extremely narrow region, with thickness of the order $2.7 \times 10^{-5}$\SI{}{m}.

\begin{figure}[H]
\centering
\subfigure[Temperature]{\includegraphics[scale=0.3]{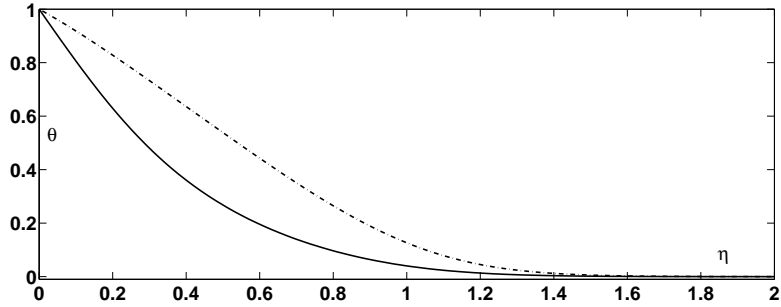}}
\hspace{0.4cm}
\subfigure[Particle concentration]{\includegraphics[scale=0.3]{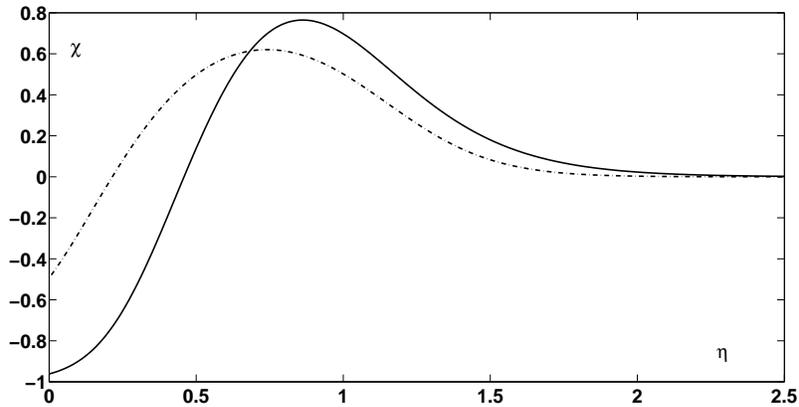}}
\caption{Comparison of a) temperatures and b) particle concentrations predicted by the current model, equations (\ref{chieq}-\ref{feq}) (solid line), and the previous system, equations (\ref{Tzouchi},\ref{Tzouth}) (dash-dotted line), using $\Sc =5, \Nb=0.1, \Nt = 0.5, \Pr=10$, $\Delta T =$ \SI{10}{K}.}
\label{Fig2}
\end{figure}

To demonstrate the effect of choosing  parameter values taken from the mathematical literature
in Figure \ref{Fig2} we compare the results of the current model, equations (\ref{chieq}-\ref{feq}), with those studied by previous authors, equations (\ref{Tzouchi},\ref{Tzouth}),  subject to the boundary conditions
(\ref{bceta1}, \ref{bceta2}). Khan and Pop \cite{Khan2010} plot results with $\Sc \in [5, 25]$, $\Nb, \Nt \in [0.1, 0.5]$, $\Pr=10$. Similar ranges are studied in the models for stretching sheet induced flow of \cite{Nogh12,Must11,Maki11,Hass11}.
In Figure \ref{Fig2} we show the temperature and volume fraction for the case $\Sc =5, \Nb=0.1, \Nt = 0.5, \Pr=10$
and again we take $\Delta T = $\SI{10}{K}.
The solid  line in the figures is the  solution of
(\ref{chieq}-\ref{feq}), dash-dot that of equations (\ref{Tzouchi},\ref{Tzouth}).
From Figure \ref{Fig2}a) we observe that the temperature predicted by the previous models is significantly higher than that of the current model. The correspondence could be worse, using different boundary conditions, but the fixed temperature condition forces both curves to pass through the same point. In this case the boundary layer extends to $\eta \approx 1.24, 1.42$ for the current and previous models.
The volume fractions are also significantly different, as seen from Figure \ref{Fig2}b).
The previous model predicts a minimum value $\chi \approx -0.5$ at the wall, which corresponds to a volume fraction
$\phi = (1+\chi) \phi_{\infty} \approx 0.5 \phi_{\infty}$. This indicates that the nanoparticles have been forced away from the wall region, they congregate in the region beyond $\eta = 0.2$ where $\phi > \phi_{\infty}$. The current model has a wall value $\chi \approx -0.98$ which is approximately twice that of the previous model. The new wall value corresponds to a dimensional volume fraction $\phi \approx 0.02 \phi_{\infty}$, in this case the particles only exceed the far field value for $\eta > 0.5$.
Importantly the use of incorrect parameter values acts to produce a significantly larger boundary layer. In Fig \ref{Fig1}b) the thickness was $\eta \approx 2.26 \times 10^{-3}$ now it is $\eta \approx 2.18$ for the current model and $1.83$ for the previous model: three orders of magnitude wider.

The values of $\phi$ are important to the discussion running through this paper. Equations (\ref{Tzouchi},\ref{Tzouth}) are derived on the assumption $\Delta \phi/\phi_{\infty} = |\phi_w -\phi_{\infty}| /\phi_{\infty} \ll 1$, yet they predict a value of 0.5. When the assumption is not made, as shown by the solid line, then $\Delta \phi/\phi_{\infty} \approx 0.98$. Making it quite clear that the assumption is invalid. In fact we should point out that errors are most likely even greater than indicated by the figure, since in both models we have neglected the variation in parameters such as density, viscosity, etc.
However, if we look back to Figure \ref{Fig1} we see that when realistic parameter values are employed then the $\phi$ variation is very small and the linearisations are permissible (but lead to rather well-known, standard equations).

\begin{figure}[H]
\centering
\subfigure{\includegraphics[scale=0.525]{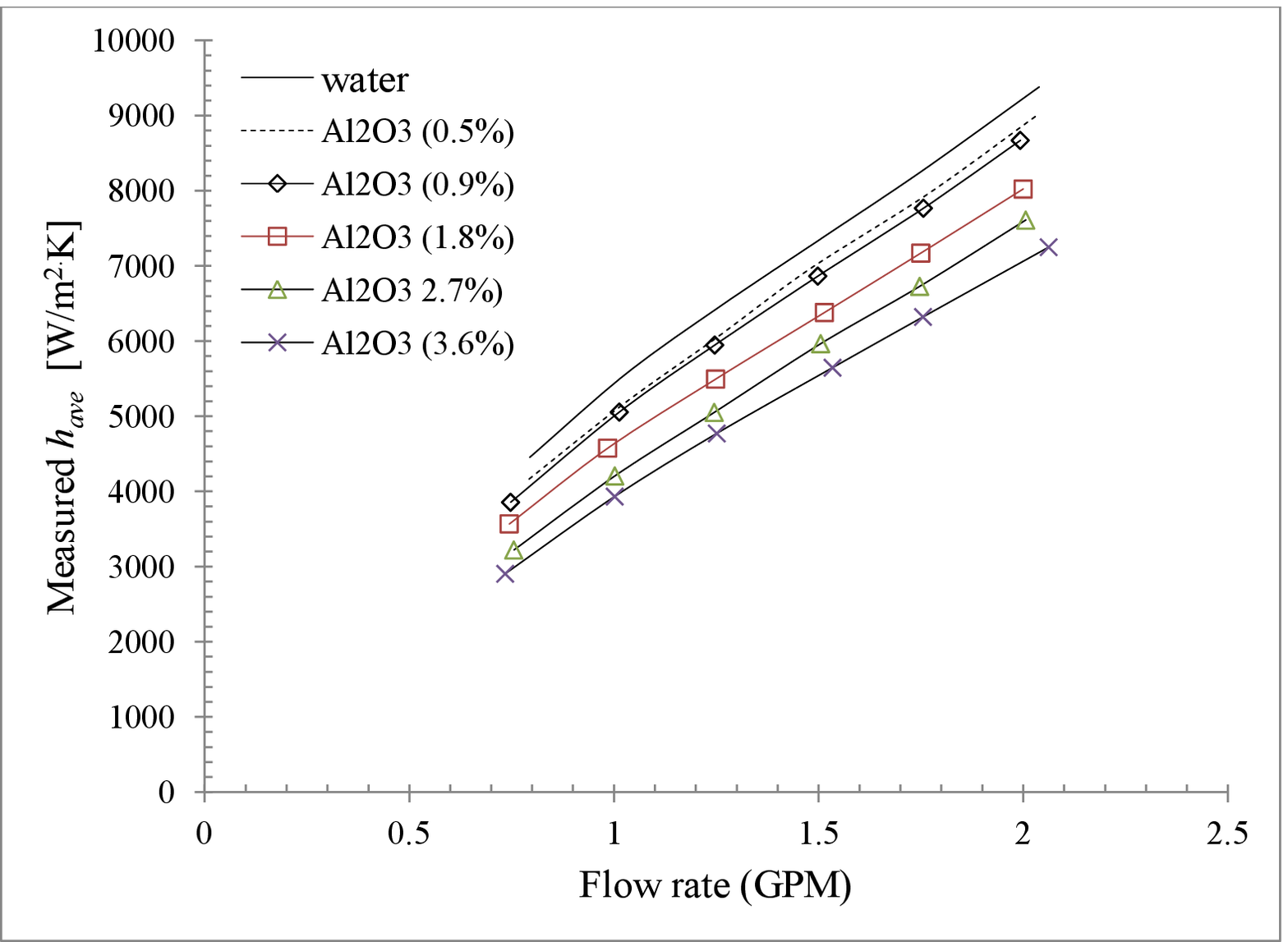}}
\hspace{0.4cm}
\subfigure{\includegraphics[scale=0.6]{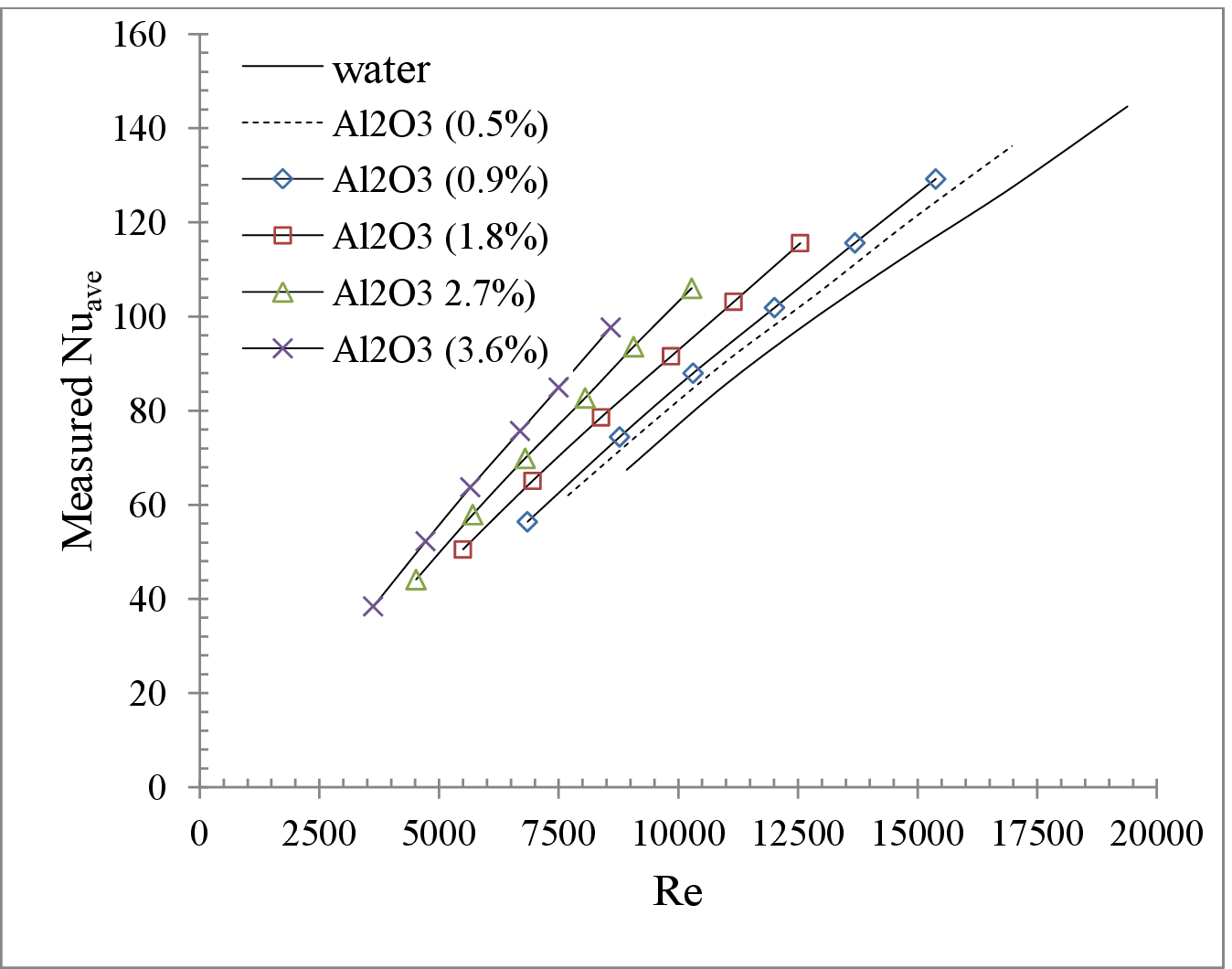}}
\caption{Comparison of a) average measured heat transfer coefficient versus flow rate and b) average measured Nusselt number versus Reynolds number. Taken from Alkasmoul \cite{Alka16}.}
\label{Fig3}
\end{figure}

The motivation in studying the boundary layer flow of a nanofluid is to determine its heat transfer characteristics and so aid in the design process. To this end the effect of flow on the Nusselt number is generally studied, where $\Nu = Q x/ (k_{nf} \Delta T) = h x/k_{nf} $, $Q$ is the heat flux at the boundary and $h$ the heat transfer coefficient. In \cite{Khan15,Khan2010,Kuzn10,Kuzn14,Maki11,Must13,Neil09b}, and many more,  results are presented in terms of the reduced Nusselt number, which depends on the Nusselt and Rayleigh or Nusselt and Reynolds number (depending on the  driving force). For the current study we require the Reynolds number, $\Rey = U x/ \nu_{nf}$. Both $\Nu$ and $\Rey$
are position-dependent and also depend on the nanofluid parameters $k_{nf}, \nu_{nf}$. In the models under discussion these parameters vary with $\phi_{\infty}$  (but in reality they should also vary with $\phi$). So any change in $\phi_{\infty}$ also changes the scaling of the non-dimensional numbers: it is therefore invalid to compare results for $\Nu$ versus $\Rey$ with different volume fraction fluids (it would make sense if the non-dimensionalisation used the base fluid values, but this is not the case).
The problem with drawing conclusions from the non-dimensional system is made clear by the results presented in Figure \ref{Fig3}. The data shown is taken from an experimental study of the heat transfer of a water-alumina nanofluid \cite{Alka16}. It was obtained by pumping nanofluids with different volume fractions through a pipe. The average heat transfer along the pipe and flow rate was recorded. The beauty of this study is that results were presented  in both dimensional and non-dimensional form.
In Figure \ref{Fig3}a) the average heat transfer coefficient is plotted against the flow rate (in gallons/minute). For a given flow rate it is quite clear that the heat transfer coefficient decreases with nanoparticle concentration. For example, when the flow rate is 1.5 gallons/minute pure water shows an average heat transfer coefficient $h \approx 7500$W/m$^2$K, whereas the nanofluid with $\phi_{\infty}=0.018$ has $h \approx 6500$W/m$^2$K. In Figure \ref{Fig3}b), where the Nusselt number is plotted against the Reynolds number, the opposite trend is observed: for a given Reynolds number the value of $\Nu$ increases  with the addition of nanoparticles, leading to the incorrect conclusion that the heat transfer has improved. The problem may be explained by
considering a specific Reynolds number, say $\Rey =10000$. From  Figure \ref{Fig3}b)  we see that $\Nu \approx 78$ for pure water, while for $\phi_{\infty}=0.018$ we find $\Nu \approx 92$.   For water $\nu = 10^{-6}$\SI{}{m^2/s} so a value $\Rey =10000$ corresponds to a velocity $U \approx$ \SI{1}{cm/s}  but when $\phi_{\infty}=0.018$ then $\nu_{nf} = \mu_{nf}/\rho_{nf} \approx 1.117 \times 10^{-6}$ \SI{}{m^2/s} and $U \approx$ \SI{1.117}{cm/s}. The graph does not make it clear that the 18\% increase in $\Nu$ was not achieved by an 18\% increase in heat transfer coefficient, it also required a 12\% increase in velocity. However, at the same time the conductivity used to define $\Nu$ also changed. It is also not clear that due to the increase in viscosity a greater pumping power/pressure drop is needed for the nanofluid. This second figure is  misleading, the dimensional plot shows the correct behaviour, which is a decrease in heat transfer with increasing volume fraction. The economics of pumping nanofluids is also discussed in \cite{Alka16}.

Note, a similar issue is discussed in \cite{Hagh15}, where they point out that the very positive results reported in the literature can be attributed to comparisons based on equal Reynolds number.
Finally the definition of the heat transfer coefficient $h = Q /\Delta T$,  used to determine $\Nu$,  does not even correctly reflect the energy transfer. The use of $\Delta T$ relies on the assumption of a linear temperature profile through the boundary layer. To correctly determine the heat transfer, and so $h$, requires integrating the temperature profile \cite{MacD14}.

\section{Conclusion}
\label{Sect:Conclusion}

In the introduction we discussed a quote from the review of Haddad \etal \cite{Hadd12} stating that the mathematical studies clearly show that nanofluids improve heat transfer in contrast to the experimental studies which show the opposite behaviour. In writing this paper we believe we have identified a sufficient number of errors and misconceptions in the mathematical studies to explain the different conclusions. Specifically we have shown that:

\begin{enumerate}

\item The non-dimensional numbers employed are typically out by a few orders of magnitude and bear no relation to the values corresponding to a physical problem.

Common previous choices correspond to `fluids' with a volume fraction greater than unity and particle sizes smaller than atoms. Quite often the Schmidt number is confused for the Lewis number.

The result of using the incorrect parameter values is to magnify the motion of the nanoparticles and so incorrectly indicate that their effect on the heat transfer is greater than in practice. On the other hand, if the correct values are employed then the problem simplifies significantly, in the case studied in this paper this reduces the volume fraction equation to a trivial form and the energy and flow equations to  forms appropriate for a Newtonian fluid.

\item The governing equations are incorrectly formulated.

If the parameter values are chosen to magnify the motion of the nanoparticles then the volume fraction and temperature can vary significantly across the boundary layer.

Density, thermal conductivity, volumetric heat capacity, viscosity, thermophoresis and Brownian motion  all  vary  with volume fraction and temperature. This dependence is neglected in the derivation of the governing equations.
Partial justification is given by stating that $\Delta T/T_{\infty} \ll 1$, which limits the application of the theory (in many practical applications the fluid can be heated to near boiling and so $\Delta T$ can be of a similar order to $T_{\infty}$). No justification is given for the linearisation of $\phi$ and, as stated above, parameter values are usually exaggerated to increase particle motion.

If the $\phi$ dependence is retained then the governing equations change and the temperature and concentration profiles also change significantly.
Ironically, if physically sensible parameter values and boundary conditions are used then the volume fraction shows very little variation and the linear approximation in $\phi$ may be justified.

\item Many models and boundary conditions describe physically unrealistic scenarios, often chosen to permit a similarity reduction. Even so the reduced models do not always satisfy the boundary conditions.

\item The comparison of results using non-dimensional numbers scaled with the nanofluid properties is misleading. If the volume fraction changes then the non-dimensionalisation also changes so, for example, an increase in Nusselt number may not represent an increase in heat transfer, it may equally well represent a decrease in nanoparticle concentration.

\end{enumerate}
While we have illustrated our analysis via the example of flow over a stretching sheet, the above conclusions hold for the many other flow configurations and boundary conditions studied in the literature.

Experiments clearly indicate that nanofluids do not provide the hoped for enhancement in heat transfer. Careful analysis of  Buongiorno's equations concur with this, but even so a mathematical industry has grown promoting the opposite result, irrespective of the experimental evidence.
In certain instances it is clear that the authors are aware of some of these issues, particularly with regard to the parameter values. In the paper that seems to have started this field \cite{Tzou08} it is stated that \lq nonlinear behaviors'  in the temperature and volume fraction solutions are noticeable for a Lewis number below 10 but in practice it is  three to four orders of magnitude greater. In \cite{Nield2009} it is stated that results diverge (from those of a standard fluid) when  $\Le$ is relatively small but tend to coincidence when
$\Le$ increases. In \cite{Nogh12} it is stated that most fluids have a large Lewis number and then quote \cite{Nield2009} to define this as $\Le > 1$. However, all these papers cite \cite{Buon06} which provides the values $\Le = 8 \times 10^5, 7 \times 10^5$ for water and alumina or copper nanofluids. In \cite{Nogh12} they also quote \cite{Buon06} for the range of diffusion coefficients  $D_B = 4 \times 10^{-12}$ to $4 \times 10^{-4}$\SI{}{m^2.s^{-1}}, yet the maximum value actually stated in  \cite{Buon06} is $4 \times 10^{-10}$\SI{}{m^2.s^{-1}}: a  difference of six orders of magnitude.

Finally, to answer the question posed in the title, \lq Does mathematics contribute to the nanofluids debate?' The answer is clearly yes, but only in that in the past it has confused and misled researchers. For example, in the conclusion of the experimental study \cite{Li10} the authors provide possible reasons for the discrepancy between theory and experiment, such as particle size distribution and interaction before suggesting an extensive comparative study involving different assumptions in the theory.

It is unlikely that the stream of theoretical research will stop based on the arguments of the present paper, however we hope that this work can at least clarify that the results obtained in the majority of the  papers quoted here (and those following on from them) do not bear any resemblance to a physical problem. Their conclusions that nanoparticles have a significant effect on heat transfer are not valid for  realistic flow configurations and only add to the confusion in the debate over whether nanofluids can work as practical coolants. It would be interesting to see if all of this endeavour could be directed in a more positive way. Nanofluids have other, more promising applications, for example in direct absorption solar cells and medicine \cite{Creg15}. Even in cooling a recent experimental study has suggested that applying a strong magnetic field to a fluid with magnetite particles can lead to a four-fold increase in heat transfer coefficient, due to the focussing of particles near the magnets, but this claim is based on a comparison of Nusselt numbers \cite{Aziz13}.
The application of mathematics to a correct set of governing equations in physically sensible parameter regimes could help in the understanding of these and other problems and so in the future benefit and guide experimental research.

\section*{Acknowledgements} V.\ Cregan and H.\ Ribera acknowledge that the research leading to these results received funding from ``la Caixa" foundation.  T.G. Myers acknowledges the support of a Ministerio de Ciencia e Innovaci\'{o}n grant MTM2014-56218. The authors thank the Institute of Thermofluids at the University of Leeds and particularly Drs Fahad Alkasmoul and Mark Wilson for granting permission to use Figure 3.

\bibliographystyle{plain}
\bibliography{nanofluids2}

\end{document}